\documentclass[conference, 9pt]{IEEEtran}

\IEEEoverridecommandlockouts
\usepackage{cite}
\usepackage{amsmath,amssymb,amsfonts}
\usepackage{algpseudocode}
\usepackage{graphicx}
\usepackage{textcomp}
\usepackage{xcolor}
\usepackage[utf8]{inputenc}
\usepackage[ruled,linesnumbered]{algorithm2e}
\graphicspath{ {Images/} }
\usepackage{listings}
\usepackage{multirow}
\usepackage{multicol}
\usepackage{array}
\usepackage{color,soul}  
\usepackage{balance}
\usepackage[backref]{hyperref}
\usepackage{booktabs}
\usepackage{url}
\usepackage{mathtools}
\usepackage{xcolor}

\usepackage{geometry}
\def\BibTeX{{\rm B\kern-.05em{\sc i\kern-.025em b}\kern-.08em
    T\kern-.1667em\lower.7ex\hbox{E}\kern-.125emX}}
\begin{document}

\setlength{\parskip}{0pt}
\title{ 
Enhancing Translation Validation of Compiler Transformations with Large Language Models
}

\author{\IEEEauthorblockN{Yanzhao Wang and Fei Xie}
\IEEEauthorblockA{\textit{Department of Computer Science} \\
\textit{Portland State University}\\
Portland, OR 97201, USA \\
\{wyanzhao, xie\}@pdx.edu}

}

\maketitle

\definecolor{keywords}{HTML}{000080}
\definecolor{functions}{HTML}{A0522D}
\definecolor{variables}{HTML}{008080}
\definecolor{parameters}{HTML}{FFA07A}

\lstdefinelanguage{myCPP}{
  language=C++,
  keywords={},
  deletekeywords={for, if}
}

\lstset{
  language=myCPP,
  basicstyle=\ttfamily\footnotesize,
  keywordstyle=[1]\color{keywords}\bfseries,
  keywordstyle=[2]\color{parameters}\bfseries,
  commentstyle=\color{gray}\itshape,
  stringstyle=\color{red},
  showstringspaces=false,
  numbers=left,
  numberstyle=\tiny\color{gray},
  numbersep=5pt,
  xleftmargin=15pt,
  morekeywords=[1]{module, func},
  morekeywords=[2]{return},
  morekeywords=[3]{},
  morekeywords=[4]{scf.yield, arith.select},
  emph={},
  emphstyle=\color{functions},
  moredelim=[s][\color{variables}]{\%}{\ },
  moredelim=[s][\color{blue}]{@}{(},
  moredelim=[s][\color{blue}]{)}{\ },
  moredelim=[s][\color{blue}]{=[ ]}{\ },
  frame=single,
  breaklines=true,
  postbreak=\mbox{\textcolor{red}{$\hookrightarrow$}\space},
}

\begin{abstract}
This paper presents a framework that integrates Large Language Models (LLMs) into translation validation, targeting LLVM compiler transformations where formal verification tools fall short. 
Our framework first utilizes existing formal verification tools for translation validation. In this work, we use Alive2, a well-known tool in LLVM compiler verification, as an example. When formal verification tools are unable to confirm a transformation's soundness, our framework employs fine-tuned LLMs for prediction. It then applies fuzzing to transformations predicted as potentially unsound by the LLMs due to return values or memory inconsistencies, aiming to find counterexamples. In cases where transformations are unsound for other reasons or sound, or if no counterexamples emerge, the framework directly reports these outcomes without further fuzzing. This methodology has shown effectiveness in complex application such as deep-learning accelerator designs, where traditional formal verification tools struggle.

\end{abstract}

\section{Introduction}

LLVM \cite{llvm}, a versatile open-source compiler, supports an extensive range of programming languages and hardware design languages. The core element of LLVM is its intermediate representation (IR). It serves as a key interface that streamlines interactions among frontends, backends, and transformation passes in the middle of the compiler, and therefore, is critical in integrating the varied components of the LLVM framework.

Translation validation, as detailed in \cite{pnueli1998translation}, has evolved into a powerful method for verifying the correctness of compiler transformations. By validating the semantics of the source and transformed IRs, it establishes an efficient and robust mechanism for ensuring the reliability of compiler transformations. Alive2 \cite{lopes2021alive2} is such a translation validation tool specialized for LLVM IR. It leverages formal verification techniques, aiming to prevent that compiler optimizations do not introduce bugs into the resultant code.

Despite their advantages, translation validation tools such as Alive2 have significant limitations. They typically cannot manage unbounded loops and have limited ability in handling external function calls. Moreover, the SMT (Satisfiability modulo theories) solver \cite{de2008z3}, a fundamental component of these tools, is often incapable of dealing with complex computations.

Our framework introduces a novel integration of fine-tuned Large Language Models (LLMs) to overcome limitations in current compiler transformation validation flows. It can conduct predictive analyses of the correctness of transformations at which formal verification tools fall short. The process begins with inputting source and target Intermediate Representations (IR), the latter being a transformed version of the former. If the verification of this transformation is deemed unsolvable by the SMT solver of formal tools such as Alive2 (the example used in this paper), it is then passed to our \texttt{LLM-based transformation predictor}. This predictor assesses the transformation, categorizing it as likely sound or unsound. For transformations predicted as unsound, the predictor also provides the reasons underlying the unsoundness. Where transformations are unsound due to return values or memory inconsistencies, our framework employs fuzzing to find counterexamples. If the \texttt{fuzzer} finds counterexamples, it confirms the transformation’s unsoundness. In cases where transformations are unsound for other reasons or sound, or if no counterexamples emerge, our framework reports their predicted soundness/unsoundness. This framework has been proven effective in practical applications, such as in Intel's deep-learning accelerator designs \cite{intelBSIM}, addressing the challenges posed by unbounded loops and complex computations.

Our primary contributions are as follows:
\begin{itemize}
    \item We pioneered the application LLMs in enhancing the reliability of compiler transformations.
    \item We seamlessly integrated Large Language Models (LLMs) with formal verification tools, exemplified by application to Alive2, to rapidly and efficiently assess LLVM transformation soundness. This integration proves particularly effective in scenarios where traditional formal verification tools struggle.
\end{itemize}

The remainder of this paper is structured as follows. Section 2 introduces the background. Section 3 presents an overview of our framework, while Section 4 dives into the specifics of our implementation. Section 5 reports our evaluation results. Section 6 discusses limitations and future work, and Section 7 concludes.

\section{Background}

Two primary methodologies are currently widely used in the domain of formal compiler verification:

1. \textbf{Theorem Proving}: This approach formally verifies if every transformation of a compiler preserves the semantics of the input program, e.g., CompCert \cite{leroy2016compcert} is a compiler for C that is formalized and verified in Coq. However, compiler certification using theorem proving is a highly complex and labor-intensive process, and every compiler revision requires reproofing. These drawbacks hinder future compiler improvements. 

2. \textbf{Translation Validation}: The translation validation approach was first introduced by \cite{pnueli1998translation}; it entails a weaker formal technique that effectively certifies the conformity of a compiler's individual executions to the compiler's specification. This approach compares a compiler's input and output for its specific application. 
\begin{figure}[ht]
\centering
\includegraphics[width=0.5\textwidth]{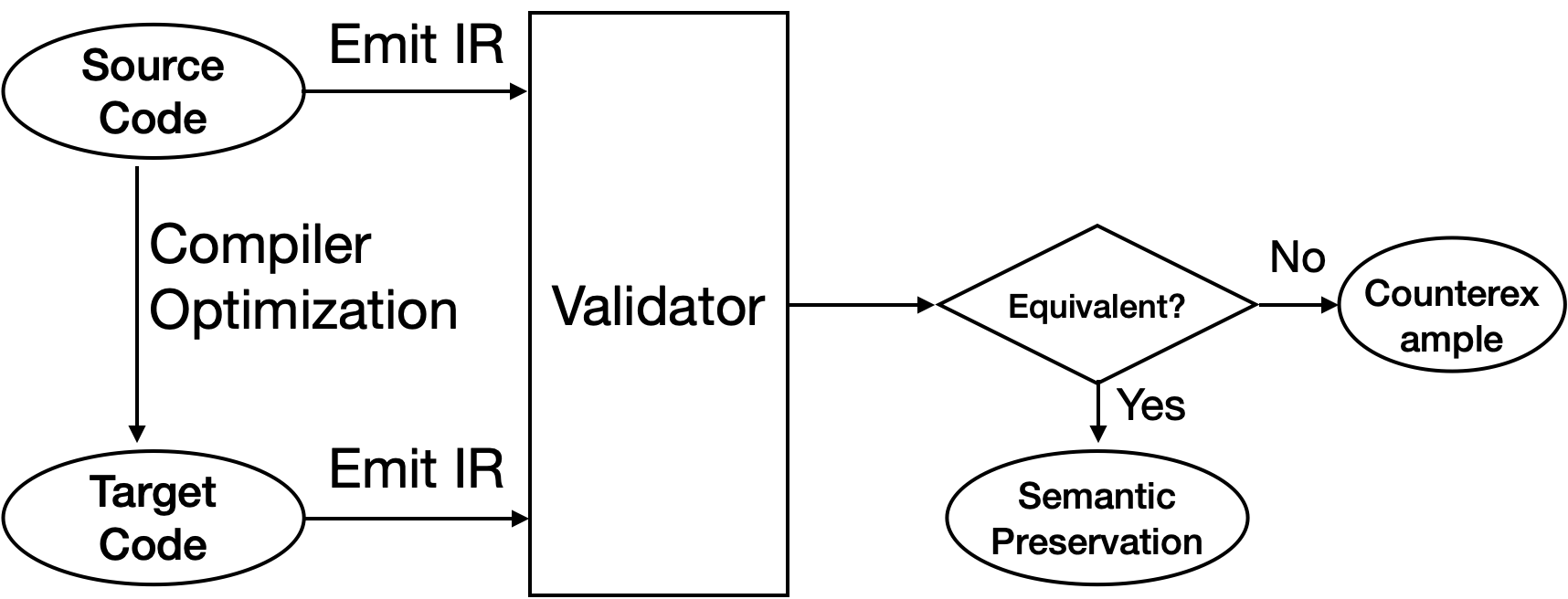}
\caption{Workflow of the Translation Validation Approach}
\label{fig:translation_validation_architecture}
\end{figure}
As illustrated in Figure \ref{fig:translation_validation_architecture}, the workflow for translation validation sends both source and target programs to a validator. If the validator confirms that the target program refines the source program, it generates a proof. Conversely, it produces a counterexample if discrepancies are detected. Translation Validation is  practical and adaptable, efficiently ensuring compiler execution aligns with specifications.

\begin{figure*}[!bht]
\centering
\includegraphics[width=1.0\textwidth]{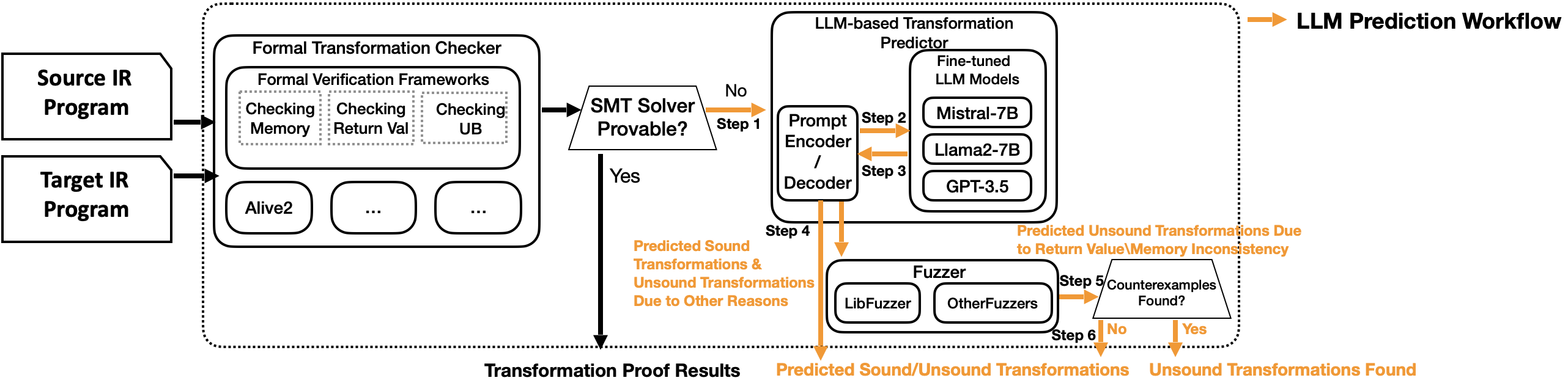}
\caption{Schematic Diagram of the Translation Validation Framework}
\label{fig:framework_workflow}
\end{figure*}

\section{Overview of The Framework}

This section presents our novel translation validation framework, consisting of three major components: the \texttt{formal transformation checker}, \texttt{LLM-based transformation predictor}, and \texttt{fuzzer}. As shown in Figure \ref{fig:framework_workflow}, the framework aims to offer an efficient approach for evaluating the soundness of compiler transformations, particularly in scenarios where traditional formal verification tools struggle.

\begin{figure}[!bht]
\centering
\includegraphics[width=0.5\textwidth]{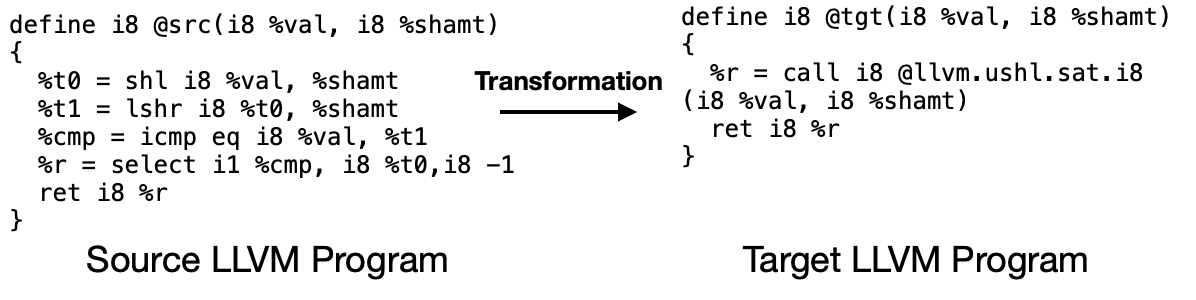}
\caption{Example of LLVM programs}
\label{fig:llvm_programs}
\end{figure}

The workflow starts with a source IR and a target IR transformed from the source IR. In this paper, we use LLVM IR as an example, as demonstrated in Figure \ref{fig:llvm_programs}. The IRs are first processed by the \texttt{formal transformation checker}, utilizing the capabilities of established translation validation tools to analyze and verify IR transformations. In this paper, we use Alive2 as the example. The \texttt{formal transformation checker} compares the memory states and return values of the transformed code, ensuring no new undefined behavior arises. This involves encoding the source and target IR programs into SMT encodings and submitting these to the SMT solver \cite{de2008z3} for verification.

If the SMT solver is able to verify the transformation, our framework reports the transformation validation results (indicating soundness or unsoundness) directly. However, if the SMT solver fails to verify the transformation, due to factors such as complex SMT expressions or unbounded loops, our framework forwards the IR program pair to the \texttt{transformation predictor} (Step 1 in Figure \ref{fig:framework_workflow}). After processing through the \texttt{prompt encode/decoder}, the input is sent to the fine-tuned LLM models, which generate the prediction results (Steps 2 and 3). Our \texttt{transformation predictor} currently supports the fine-tuned models: Mistral-7B \cite{jiang2023mistral}, Llama2-7B \cite{touvron2023llama}, and GPT-3.5 \cite{floridi2020gpt}. The fine-tuning process of these LLM models is elaborated in a subsequent section.

Then, for transformations predicted as unsound due to inconsistent memory or return values, our framework sends both the program pair and the reasons for unsoundness to the \texttt{fuzzer} to identify counterexamples (Step 4). If the \texttt{fuzzer} finds any counterexamples, the transformation is deemed unsound (Step 5 and 6). For transformations predicted as sound, and for unsound transformations where no counterexamples are found or those caused by other reasons, our framework directly reports their predicted soundness/unsoundness (Step 4).

\section{Implementations}

\subsection{Formal Transformation Checker}

The formal transformation checker is crucial for analyzing compiler transformations. It focuses on verifying memory states, return values, and preventing the introduction of new undefined behaviors (UBs), thereby ensuring transformation correctness and reliability.

\subsubsection{Checking Memory States}

The checker uses existing formal verification tools for a detailed comparison of memory states between the source IR \(P_{\text{src}}\) and the target IR \(P_{\text{tgt}}\). It verifies that memory operations in \(P_{\text{src}}\) and \(P_{\text{tgt}}\) are equivalent, denoted as:

\[
\forall \sigma_{\text{src}}, \sigma_{\text{tgt}} : \text{Memory}(P_{\text{src}}, \sigma_{\text{src}}) \equiv \text{Memory}(P_{\text{tgt}}, \sigma_{\text{tgt}})
\]

\subsubsection{Checking Return Values}

Similar to memory state verification, the checker ensures consistent return values post-transformation, preserving the program's functional behavior. It encodes return values from both source and target programs and employs a SMT solver, such as Z3 \cite{de2008z3}, to compare these SMT encodings, flagging any discrepancies.

\subsubsection{Detecting Introduction of New Undefined Behaviors}

The checker detects new undefined behaviors (UBs) in the transformed program, ensuring no new UBs are introduced. This involves checking against division by zero, null pointer dereferencing, and other unpredictable or unsafe operations, thereby maintaining the program's correctness and reliability.

\subsection{LLM-Based Transformation Predictor}

This subsection introduces our primary innovation: fine-tuning Large Language Models (LLMs) in enhancing the robustness of the compiler transformations.

Currently, our training data originates from the verification outcomes of llvm-project \cite{lattner2004llvm} transformations and unit tests conducted by Alive2, while similar data from other tools could also be applied. After eliminating duplicate records, our dataset comprised 32,850 sound transformations and 405 unsound transformations. We randomly selected 40 pairs from these to form our test dataset and allocated the remainder to the training set. For each unsound transformation, we included reasons for the unsoundness, such as memory issues, return values, or the introduction of new undefined behaviors (UB). We then formatted this data into prompts for LLM fine-tuning, utilizing the \texttt{prompt encode/decoder} within the \texttt{transformation predictor}. We defined the fine-tuning prompt structure as follows:

\begin{itemize}
    \item \textbf{System Content:} "Your task involves analyzing the given IR transformations. On receiving a transformation in the "Transformation: \texttt{X} $\rightarrow$ \texttt{Y}" format, analyze and respond in the "Status: \texttt{A} Reason: \texttt{B}" format. "\texttt{A}" denotes the transformation's correctness as CORRECT or UNSOUND. For UNSOUND transformations, list reasons "\texttt{B}". Your analysis involves a two-step approach:
\begin{itemize}
\item \textbf{Special Value Injection}: Inject specific values into both original and transformed IR to observe and compare behaviors.

\item \textbf{Step-by-Step Computation and Analysis}: Execute detailed computations for both IR versions to pinpoint discrepancies.
\end{itemize}
Base your analysis on the following to assess soundness:
\begin{itemize}
    \item 
\textbf{Undefined Behavior Consistency}: The target should only trigger UB if the source does. New UB in the target renders the transformation unsound.
    \item 
\textbf{Return Domain Consistency}: The target's return domain must align with the source's, except when the source triggers UB. A mismatched return domain without source UB suggests unsoundness.
    \item 
\textbf{Poison Value Propagation}: The target's return value should indicate poison only if the source’s does. Any additional poison in the target signals unsoundness.
    \item 
\textbf{Undefined Value Handling}: The target's return value should be Undefined only if the source’s is Undefined or poison. Introduction of Undefined values by the target without source justification is unsound.
    \item 
\textbf{Return Value Consistency}: The return values of both the source and target should match when the source is clear of Undefined or poison. Variances under a well-defined source indicate unsoundness.
    \item 
\textbf{Memory State Refinement}: Verify that the memory state after target execution refines that of the source's. Memory state inconsistencies suggest unsoundness.
\end{itemize}

    \item \textbf{User Content:} ``\{\textit{Source IR}\} $\rightarrow$ \{\textit{Target IR}\}''
    
    \item \textbf{Assistant Content:} "Transformation status: \{\textit{SOUND/UNSOUND}\} Reason: \{\textit{UNSOUND REASON}\}"
\end{itemize}

Choosing the appropriate training data is crucial. Our dataset is imbalanced, with far more sound than unsound transformations. Using the dataset directly could make the model biased. To alleviate the potential bias issue, for sound transformations, we removed 1,874 instances where the source and target code were exactly the same. Then we sampled the sound and unsound data at ratios of 1:1, 2:1, 4:1, and 8:1 to evaluate the performance.

After fine-tuning, when our framework identifies transformations that the SMT solver cannot prove, it forwards the transformation pair to the \texttt{transformation predictor}. First, this data is structured using the \texttt{prompt encoder}, adhering to the fine-tuning prompt structure. Next, the formatted data is sent to the LLMs. The LLMs then predict the soundness/unsoundness of the transformations. Finally, the \texttt{prompt decoder} converts the models' output into a human-readable text format, completing the transformation prediction process.

\subsection{Fuzzer}

The \texttt{fuzzer} module is a key part of our framework. It aims to find concrete counterexamples that confirm the unsoundness due to inconsistent memory or return values predicted by the \texttt{transformation predictor}. For LLVM transformations, our framework uses LibFuzzer \cite{lattner2004llvm}.

The \texttt{fuzzer} takes program pairs and their LLM-predicted unsound reasons given by the framework as inputs. The input format is:

\[
\text{Input: } \{ \text{Source IR}, \allowbreak \text{Target IR}, \allowbreak \text{Unsoundness Reasons} \}
\]

Based on the predicted reasons for unsoundness, the \texttt{fuzzer} employs different strategies to fuzz the predictions:

\begin{itemize}
\item \textbf{Handling Return Value Inconsistencies:} If the unsoundness is caused by return value issues, the \texttt{fuzzer} focuses on the return logic of the program pair. It tests both programs under various inputs to spot any output differences.
\item \textbf{Handling Memory Inconsistencies:} For unsoundness caused by memory, the \texttt{fuzzer} examines memory operations in both source and target IRs. It aims to find any differences in memory state after execution, checking that the target IR does not introduce new memory behaviors or access issues not in the source IR.
\end{itemize}
\section{Evaluation Results}

\begin{figure*}[ht]
\centering
\includegraphics[width=0.75\textwidth]{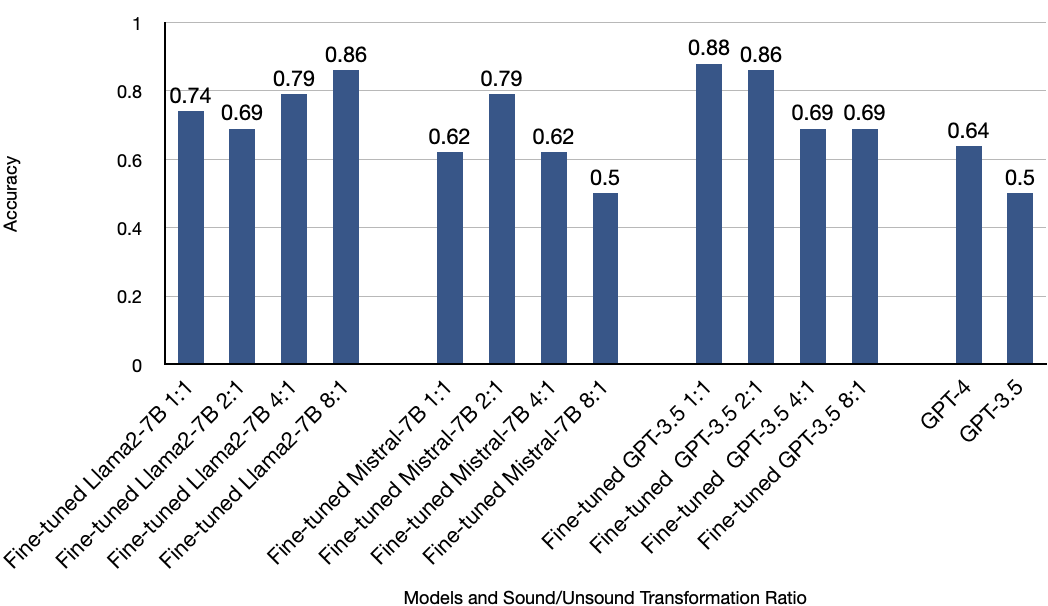}
\caption{Evaluation Results on LLVM Transformations}
\label{fig:evaluation_results}
\end{figure*}

In this section, we present the evaluation results of our framework, focusing on assessing the accuracy of LLMs in determining the soundness of compiler transformations. We selected 80 transformations from the \texttt{llvm-project} and \texttt{Alive2}'s unit tests, maintaining a balanced ratio of sound to unsound results. For locally deployed models such as \texttt{Llama2-7B} and \texttt{Mistral-7B}, we conducted our runs on an Apple M3 Max 128GB platform. For \texttt{GPT-3.5}, we accessed the model via the OpenAI API. Additionally, we assessed the baseline performance of the original \texttt{GPT-4} and \texttt{GPT-3.5} models without fine-tuning.

We conducted comparison of the fine-tuned \texttt{Llama2-7B}, \texttt{Mistral-7B}, and \texttt{GPT-3.5} models, as well as the \texttt{GPT-4} and \texttt{GPT-3.5} models without fine-tuning, using the same test data. These models were evaluated based on their accuracy, considering various ratios of SOUND to UNSOUND data in the training set, specifically 1:1, 2:1, 4:1, and 8:1.

As depicted in Figure \ref{fig:evaluation_results}, the fine-tuned \texttt{GPT-3.5} model achieved an accuracy of up to 88\%, while locally deployable models, like the fine-tuned \texttt{Llama2-7B}, reached an accuracy of up to 86\%, and the \texttt{Mistral-7B} reached an accuracy of up to 79\%. The \texttt{GPT-4} model, without fine-tuning, attained only an accuracy of 64\%, and the \texttt{GPT-3.5} without fine-tuning achieved only 50\% accuracy. These results demonstrate that LLMs, when fine-tuned with an appropriate dataset, can effectively provide rapid assessments of compiler transformations. Moreover, it shows that properly fine-tuned smaller models, such as \texttt{Llama2-7B} and \texttt{GPT-3.5}, can outperform larger models like \texttt{GPT-4}. This evaluation highlights the potential of LLMs in assessing the soundness of compiler transformations.

\subsection{Evaluation of Deep-learning Accelerator Designs in LLVM IR}

\begin{figure}[ht]
\centering
\includegraphics[width=0.5\textwidth]{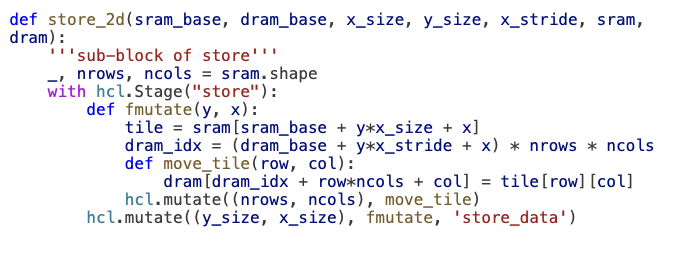}
\caption{load\_2d module from hVTA}
\label{fig:load_2d}
\end{figure}

We employed our framework to assess Intel deep-learning accelerator designs \cite{intelBSIM} specified using HeteroCL \cite{lai2019heterocl}, which features a Python-based domain-specific language (DSL) for specifying hardware designs at a high abstraction level. It then compiles the hardware from this DSL to a lower-level design language such as LLVM, and that of high-level synthesis. In this evaluation, we utilized one source deep-learning design (named hVTA), and two transformed designs from the source design (named sVTA and uVTA) as examples. These designs adhere to the VTA open-source accelerator specifications\footnote{\small \url{github.com/apache/tvm/blob/v0.6/vta/include/vta/hw_spec.h}} and consist of four modules: \texttt{load}, \texttt{ALU}, \texttt{GEMM}, and \texttt{store}. 

\begin{figure*}[ht]
\centering
\includegraphics[width=0.95\textwidth]{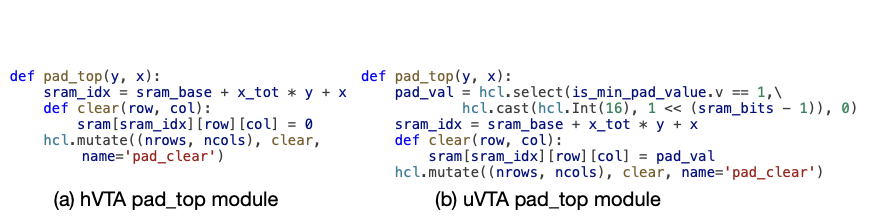}
\caption{load\_2d module from hVTA-uVTA}
\label{fig:pad_top}
\end{figure*}

We discovered that individual modules generated LLVM code exceeding the LLMs' 4096 token limit. Therefore, we further divided the modules into smaller, functionally discrete units such as store\_2d, load\_data, load\_uop, and others, totaling 19 modules. Figure \ref{fig:load_2d} shows an example of the store\_2d module's code from hVTA, illustrating how it performs tiling according to VTA's SRAM shape and writes data into DRAM. Out of these 19 modules, 12 were not successfully processed by Alive2. For each of these 12 modules from hVTA, sVTA, and uVTA, we used hVTA's modules' LLVM IR as the source program and the respective modules' LLVM IR from sVTA and uVTA as the targets. We employed a fine-tuned GPT-3.5 model, with a SOUND: UNSOUND data ratio of 2:1, to predict the soundness of their LLVM IR. The model indicated UNSOUND results for the pad\_top modules in hVTA and uVTA with inconsistent memory. The fuzzer generated a counterexample with \texttt{is\_min\_pad\_value} set to \texttt{true}. As shown in Figure \ref{fig:pad_top}, we found that uVTA altered hVTA's behavior. While the hVTA module directly wrote 0 to SRAM, uVTA's value written to SRAM depended on pad\_val.

\section{Limitations and Future Work}

This section discusses current limitations of our approach and outlines potential directions for future work.

\textbf{Limitations:}
Our framework, despite promising results, has limitations. One primary constraint is the limited diversity of our dataset. We managed to collect only 405 unsound transformations, which may not sufficiently represent the wide spectrum of potential unsound transformations in the wild. This limitation could affect the generalizability of our model's predictions.

Moreover, our fine-tuning efforts were constrained by platform capabilities. We only conducted the fine-tuning of the Llama2-7B model and Mistral-7B locally. Larger models may yield better performance.

Additionally, the inherent limitations of the models' context window restrict our framework's ability to handle larger programs. This limitation can prevent the processing of more extensive and complex program structures, posing a challenge to the scope of transformations that our framework can accurately evaluate.

\textbf{Future Work:}
To address these limitations and enhance the robustness of our framework, we plan to undertake the following initiatives in our future work:
\begin{itemize}
    \item \textbf{Synthetic Data Generation}: We are considering developing rules for synthetic data generation. This approach could potentially enable us to generate a larger number of training samples, thereby enhancing the training process and potentially improving the model's predictive accuracy. A larger and more diverse dataset will likely improve the model's ability to generalize and accurately predict the correctness of LLVM transformations.
    
    \item \textbf{Exploring Larger Models}: We intend to experiment with fine-tuning larger LLM models. Larger models have a greater capacity for learning and might demonstrate superior performance in predicting transformation correctness.
    
    \item \textbf{Exploring Other IRs}: While our method is currently using LLVM IR and Alive2 as examples, it is not inherently limited to them. In our future work, we also aim to explore its applicability to other IRs, such as MLIR \cite{lattner2021mlir}, broadening the scope and versatility of our framework.

\end{itemize}


\section{Summary}

In this paper, we have presented a Translation Validation framework integrated with LLMs. We utilized verification data from an existing formal verification tool's processing of LLVM transformations as the training set to fine-tune models such as Llama2-7B, Mistral-7B, and GPT-3.5. This approach aimed to provide a predictive mechanism for transformations that prove challenging for traditional formal verification tools. The evaluation results underscore the potential of LLMs in enhancing the robustness of compiler transformations.


\begingroup
\footnotesize
\bibliographystyle{abbrv}
\bibliography{ref}
\endgroup

\end{document}